\documentclass[preprint,aps,pre,showpacs]{revtex4}
\usepackage{epsf, epsfig, graphics,amsmath}

\begin{document}
\draft
\title{Determine the strength of soft bonds}
\author{Hsuan-Yi Chen and Yi-Ping Chu}
\affiliation{Department of Physics and Center for Complex Systems,
National Central University, Chungli, 32054  Taiwan}
\date{\today}
\begin{abstract}
The strength of a simple soft bond under constant loading rate is studied 
theoretically.  We find 
a scaling regime where rebinding is negligible and the rupture 
force of the bond scales as 
$const. + (\ln (kv))^{2/3}$, where $kv$ is the loading rate. 
The scaling regime is smaller for weaker bonds and broader for stronger bonds.
For loading rate beyond the upper limit of the scaling regime, bond 
rupture is deterministic and thermal effects are negligible.  
For loading rate below the lower limit of the scaling regime, 
contribution from rebinding becomes important, and there is no simple 
scaling relation between rupture force and loading rate.
When we extend the theory to include the effect of rebinding 
we find good agreement between theory and simulation even below the scaling
regime.
\end{abstract}
\maketitle

Non-covalent molecular bonds in biological systems are soft.
Typical binding energy 
is on the order of $10 k_BT$.  This extremely weak binding
energy enables fast response to external stimuli which is preferred by living 
systems.~\cite{ref:Lodish_book}  
At the same time, these weak bonds are sensitive to thermal 
fluctuations, thus the study of the mechanical response of the bonds has to
consider thermal effects.  
Experimentally, the strength of a soft bond is often characterized by 
the relation between the rupture force and loading rate, i.e., 
the dynamic force spectroscopy (DFS).~\cite{ref:Evans_ARBBS_01}
The imposed loading rate in an experiment provides a time 
scale that can be used to probe the internal dynamics of the bond under 
study.  Therefore DFS experiments of soft bonds have revealed valuable 
structural and dynamical information for 
biological soft materials ranging from the energy landscape of several
non-covalent bonds to the unfolding mechanisms of several types of 
proteins~\cite{ref:Evans_ARBBS_01}.

The theoretical problem of finding the rupture force of a bond under
constant loading rate corresponds to evaluating the escape time of a 
particle in a time-varying one-dimensional
potential well in the presence of thermal fluctuations.  
This is an interesting extension of the celebrated 
Kramers escape theory~\cite{ref:Kramers_physica_40,ref:RMP_90,ref:Garg_PRB_95}.
To study experimentally relevant problems, extensions of the theory to 
systems with multiple bonds in parallel~\cite{ref:Hummer_BPJ_03} or 
in series~\cite{ref:Seifert_EPL_02}, and extensions of 
the theory to bonds which are described by higher dimensional 
potentials~\cite{ref:Ajdari_PRE_02,ref:Ajdari_BPJ_04} have been studied 
recently.  In the case of multiple bonds, experiments~\cite{ref:Merkel_PRL_02} 
have found good qualitative agreement with 
theories~\cite{ref:Seifert_EPL_02,ref:Hummer_BPJ_03}.  
In the case of bonds with more than one energy minimum, the
theory~\cite{ref:Ajdari_BPJ_04} 
provides new interpretation for experimental results.  

Despite the successes on both experimental and theoretical sides, it is
surprising that the conventional ``linear
theory''~\cite{ref:Bell_78, ref:Evans_ARBBS_01, ref:Evans_BPJ_97} for the
DFS of a simple one-dimensional bond, the corner stone of theoretical 
analysis of most experimental studies, is challenged recently by two different 
theories that we refer to as harmonic theory~\cite{ref:Hummer_BPJ_03} and
cubic theory~\cite{ref:Dudko_PNAS_03} respectively.
Linear theory assumes that under constant loading rate the activation barrier 
for bond dissociation diminishes linear in time.  The
resulting rupture force scales like $\ln (kv)$ where $kv$ is the 
loading rate.
Recently, Hummer and Szabo~\cite{ref:Hummer_BPJ_03} treated the
bond as a harmonic potential but ignores 
rebinding and details of potential shape near the barrier, thus
we call this theory harmonic theory.  
The harmonic theory predicts a relation between the rupture force and the
loading rate with the form 
$const. + (\ln kv)^{1/2}$.  At about the same time,
Dudko et. al.\cite{ref:Dudko_PNAS_03} assumed that bond rupture occurs when the
potential energy near the barrier can be approximated by a cubic function, 
thus we call this theory cubic theory.
Cubic theory follows an earlier work by Garg~\cite{ref:Garg_PRB_95},neglects
rebinding, and shows that the rupture force scales like 
$const. + (\ln kv)^{2/3}$.
Ref.~\cite{ref:Hummer_BPJ_03} showed that the harmonic theory can be extended 
to describe the unfolding of protein titin, 
and Ref.~\cite{ref:Dudko_PNAS_03} 
showed that the scaling prediction in cubic theory for Morse potential is
better than linear theory by comparing the fitting of rupture force in 
simulation to $(\ln (kv))^{2/3}$ and $\ln (kv)$ respectively.
Both Ref.~\cite{ref:Hummer_BPJ_03} and~\cite{ref:Dudko_PNAS_03} 
did not provide a quantitative analysis for the effect of rebinding.  
In another seemingly unrelated field of the physics of  
the atomic scale friction, linear theory~\cite{ref:Gnecco_PRL_00} and 
cubic theory~\cite{ref:Grant_PRL_01} have both been 
proposed to describe the rupture of bonds (with no rebinding) between two 
parallel surfaces under relative motion.  Although data from friction
experiment~\cite{ref:Sills_PRL_03} is not able 
to discriminate between linear and cubic theory, it is believed that for
atomic friction linear
theory works for weak bonds with binding energy slightly greater than $k_BT$ 
and cubic theory is better for bonds with realistic binding energies which
is on the order of $10k_BT$~\cite{ref:Grant_PRL_01,ref:Filippov_PRL_04}.

To discriminate between linear, harmonic, and cubic theory, and to provide
the underlying physical picture, including rebinding,
for the rupture of molecular bonds under
constant loading rate, in this Letter we present an analysis of simple bond 
DFS. Unlike Ref.~\cite{ref:Hummer_BPJ_03,ref:Dudko_PNAS_03}, we do not discuss 
higher dimensional problems.  We focus on simple bonds, we consider loading
rate beyond the scaling regime, and we consider the effect of rebinding in
our theory.  Our result shows that an {\em extended cubic theory} which takes 
rebinding into account is suitable to describe the rupture of simple bonds
even at low loading rate beyond scaling regime. 

In a typical DFS experiment, a bond attached to a surface on one end and 
a spring on the other end is pulled out by the spring with
a constant velocity $v$ until the
bond breaks and the spring recoils to its rest position. 
For a sufficiently simple system, a single reaction
coordinate $x$, usually the distance between the particle and the
substrate, is sufficient to describe the dynamics of the system. 
Let $U_0(x)$ be the potential of the bond and $L_0$ be its equilibrium 
position in the absence of external forces.
The potential of the spring is 
$U_{spring}(x,t) = \frac{k}{2}(vt+L_0-x)^2$ for $t \ge 0$, 
where $k$ is the spring constant.
When the total potential $U(x,t)= U_0(x)+U_{spring}(x,t)$ is bistable, 
the minimum at smaller $x$ is due to the bond, and it is 
denoted as $x_a(t)$,
the minimum at larger $x$ is due to the spring, and it is 
denoted as $x_f(t)$,
the barrier is denoted as $x_b(t)$.
$x_a(t)$, $x_b(t)$, and $x_f(t)$ are determined by solving
$-kvt=-(d U_0/d x) - k(x-L_0)\equiv \tilde{F}(x)$. 
As Figure 1 shows, the solutions to $-kvt = \tilde{F}(x)$ at time
$t$ are the crossing points between the stationary curve $y=\tilde{F}(x)$ 
and the moving horizontal line $y=-kvt$ which scans downward with constant
rate $kv$.  
Between $t_1$ and $t_2$, $-kvt=\tilde{F}(x)$ has three solutions
and $U(x,t)$ is bistable.
When $t<t_1$ the bond is stable, 
when $t>t_2$ the bond is unstable. 
We further denote the local minimum (maximum) of $\tilde{F}(x)$ as 
$x_1$ ($x_2$),i.e., $-kvt_{1(2)} = \tilde{F}(x_{2(1)})$.  

The dynamics of the system along the reaction coordinate is
described by the Langevin equation,
\begin{eqnarray}
\frac{dx}{dt} = - \gamma \frac{\partial U(x,t)}{\partial x} +
\zeta (t).
\end{eqnarray}
Here $\gamma$ is the damping coefficient, $\zeta$ is the thermal
noise with zero mean and variance $2k_BT \gamma$. 
For simplicity, we assume $\gamma$ to be independent of $x$.
The energy unit is chosen to be $k_BT$, $L_0$ is unit length, and 
unit time is chosen such that $\gamma =1$.
When $U(x,t)$ is bistable, the system is in the bound state
if $x(t) < x_b(t)$, otherwise it is in the free state. 
The probability $\phi (t)$ that the system is in the 
bound state at time $t$ satisfies the reaction equation 
\begin{eqnarray}
\frac{\partial \phi (t)}{\partial t} = -k_{off}(t) \phi (t) 
                                      + k_{on}(t) (1-\phi (t)),
\end{eqnarray}
the probability that bond rupture occurs at time $t$ is
\begin{eqnarray}
 P_b(t) = \phi (t) k_{off}(t) e^{-\int _t^{t_2} dt' k_{on}(t')}
        = \left(
               1- \int _{t_1}^t dt' k_{off}(t')
                       e^{-\int_{t'}^t (k_{on}(t'') + k_{off}(t'') )dt''} 
          \right) k_{off}(t) e^{-\int _t^{t_2}k_{on}(t')dt'}
\label{eq:rupturetime}
\end{eqnarray}  
and the mean rupture force is 
\begin{eqnarray}
 \langle k(vt+L_0-x_a(t)) \rangle 
= \int _{t_1}^{t_2} dt P_b(t) k(vt+L_0-x_a(t))
\label{eq:ruptureforce}
\end{eqnarray} 
It is reasonable to assume that time scales to reach local
equilibrium in both bound and free states are much smaller than
time scales of transitions between the two states.   
Thus Kramers escape theory~\cite{ref:RMP_90,ref:Kramers_physica_40} 
states that the on- and off-rates satisfy  
\begin{eqnarray}
 k(t) = \frac{1}{2\pi}
\sqrt{ U''(x_{\alpha},t) | U''(x_{b},t) |} e^{-(U(x_{b} - U(x_{\alpha}))},
\end{eqnarray}
where $U''(x,t)=\partial^2U(x,t)/\partial x^2$. 
For $k(t)=k_{off}(t)$, $x_{\alpha} = x_a(t)$; for $k(t) = k_{on}$, 
$x_{\alpha} = x_f(t)$ 

Because bond rupture should occur at $t$ closer to $t_2$ than $t_1$, 
from Fig.~1 it is natural to approximate $\tilde{F}(x)$ 
by a quadratic function near $x_1$, and by a straight line near $x_f(t)$.  
The resulting approximation for $\tilde{F}(x)$ is
\begin{eqnarray}
  \tilde{F}(x) \approx \left\{ \begin{array}{ll}
                      \frac{c}{2} (x-x_1)^2 + f_0, & x < x _2  \\
                      -k(x-L_0), & x > x_2.
                    \end{array} \right.
\label{eq:fp}
\end{eqnarray}
Here $c = - \left( \frac{d^3 U_0}{dx^3} \right)_{x=x_1}$,
and $f_0 = - \left( \frac{d U_0}{d x} \right) _{x=x_1} - k (x_1-L_0)$.
Eq.~(\ref{eq:fp}) corresponds to a potential that is
cubic in $x$ near $x_1$, and is the spring potential near $x=L_0+vt$.  
That is, cubic theory~\cite{ref:Dudko_PNAS_03}, not linear or harmonic
theory, follows naturally from 
observing the generic shape of $\tilde{F}(x)$.  
A straightforward calculation
leads to Kramers on- and off-rates in the cubic theory.    
\begin{eqnarray}
k_{on}(t) &=& \frac{1}{2\pi} \sqrt{ k \left(2c (-f_0-kvt) \right)^{1/2} }
             \exp \left\{ -\frac{c}{3}  \left( \frac{2}{c} (-f_0-kvt)
             \right)^{2/3}-\frac{k}{2} (x_1-L_0-vt)^2 -U_0(x_1)
             \right\}, \nonumber \\
k_{off}(t) &=& \frac{1}{2 \pi} \sqrt{2c (-f_0-kvt)} \exp
  \left\{ -\frac{2c}{3} \left( \frac{2}{c} (-f_0-kvt) \right)^{2/3}
  \right\}.
\label{eq:rates}
\end{eqnarray}

At sufficiently large $kv$ rebinding is unlikely and 
\begin{eqnarray}
   P_b(t) &=& k_{off}(t) \times  \exp \left( \int _{t_1}^{t} \ k_{off}(t')
   dt'\right)  \nonumber \\
   &\approx & \frac{\sqrt{-2c (f_0 + kvt)}}{2 \pi} \exp \left\{
   -\sqrt{ \frac{32}{9c}} (-f_0-kvt)^{3/2} - \frac{c}{4\pi kv}
   e^{-\sqrt{ \frac{32}{9c}} (-f_0-kvt)^{3/2}}\right\},
\label{eq:pb}
\end{eqnarray}
and the rupture force then has the simple form
\begin{eqnarray}
   F_{max} = \left( \frac{dU_0}{dx} \right)_{x=x_1} + k  \sqrt{
   \frac{2\left( \sqrt{\frac{9c}{32}} \ln \frac{c}{4\pi kv} \right)^{2/3} }{c}} - \left( \sqrt{\frac{9c}{32}} \ln \frac{c}{4\pi kv} \right)^{2/3}.
\label{eq:fmax}
\end{eqnarray}
The second term on the right hand side of Eq.~(\ref{eq:fmax}) is 
$k \langle x_1 - x_a(t) \rangle$, in most cases it is negligible, thus cubic
theory predicts the scaling relation $F_{max} \sim const. + (\ln (kv))^{2/3}$.
This scaling relation has been tested by data collapsing for simulating
Morse potential in Ref.~\cite{ref:Dudko_PNAS_03}.  Here we provide direct
comparison between simulation and theory for a couple of potentials so that
any discrepancy between theory and simulation can be examined critically.
Figure 2 compares the rupture force predicted by Eq.~(\ref{eq:fmax})
and numerical simulations for Morse potential 
$U_0(x)=W ([1-\exp (-2b(x-L_0)/L_0)]^2-1)$ with $b=1.5$ and a power law 
potential $U_0(x) = W(\frac{1}{2(x/L_0)^6}-\frac{3}{2(x/L_0)^2})$ 
which is taken from Ref.~\cite{ref:Seifert_EPL_02}, the spring constant
is chosen to be $k=0.8$.
Our result shows better agreement between Eq.~(\ref{eq:fmax}) and simulations
for stronger bonds and deviation of simulation from cubic theory at high and 
low loading rates.

The deviation of simulation from Eq.~(\ref{eq:fmax}) at high loading rate
can be understood by noticing that according Eq.~(\ref{eq:pb}), 
the probability that bond rupture occurs between $t_1$ and $t_2$ is 
\begin{eqnarray}
    \int _{t_1}^{t_2} P_b(t) dt \approx 1-e^{-\frac{c}{4\pi kv}},
\end{eqnarray}
i.e, at high loading rate with $c/4\pi kv \leq 1$, bond rupture 
is not likely to occur between $t_1$ and $t_2$.  
This is because from $t$ to $t_2$ $x_a(t)$ moves a distance 
$x_1-x_a(t)= \sqrt{ 2kv(t_2-t)/c}$, and 
the displacement of a particle under free diffusion during the same time
interval is $x_{d}(t)= \sqrt{2(t_2-t)}$, thus
$c/4 \pi kv \sim (x_d/(x_1-x_a(t)))^2$ compares free diffusion
and the motion of $x_a(t)$ due to the spring. When $c/4\pi kv \geq 1$,
i.e., $x_1-x_a(t) <x_{d}(t)$, free diffusion near $x_a(t)$ is faster than the 
motion of $x_a(t)$, therefore the particle senses the barrier and bond rupture
is due to barrier crossing, and Kramers theory is appropriate to describe bond
rupture.  On the other hand, if $x_1-x_a(t) > x_{d}$, free diffusion alone 
cannot catch up the motion of potential minimum, therefore the particle is
located near $x_a(t)$ at all time until the barrier vanishes.  That is, 
thermal effect is negligible in this case.  The filled symbols in the high $v$ 
region of Figure 2 shows where $c/4 \pi kv >2$, and that is when 
Eq.~(\ref{eq:fmax}) starts to deviate from simulation result.  

The deviation of simulation from Eq.~(\ref{eq:fmax}) at low loading rate is
due to the effect of rebinding.  When the chance of observing a 
rebinding event between the mean rupture time $\langle t_r \rangle $ 
predicted by Eq.~(\ref{eq:pb}) and $t_2$ is not negligible, i.e.,
\begin{eqnarray}
 && \int _{\langle t_r \rangle }^{t_2} k_{on}(t) dt \nonumber \\
&=&   
  \frac{c}{4\pi kv} \left( \frac{8k^2}{c} \right)^{1/4} e^{-U_0(x_1)}
    \int _{\left( \frac{c}{4\pi kv} \right)^{-1/2}} ^1 dw
   \left( -\sqrt{ \frac{9c}{8}}\ln w \right)^{-1/6}
   e^{ -\frac{1}{2k} \left( \left( -\sqrt{\frac{9c}{8}}
   \ln w \right)^{2/3} - U_0'(x_1) \right)^2 } \nonumber \\
&\geq& \mathcal{O}(1),
\end{eqnarray}
rebinding has to be included in calculation for $F_{max}$.
Filled symbols in the low $v$ region of Fig.~3 are where 
$\int _{\langle t_r \rangle }^{t_2} k_{on}(t) dt \geq 0.1$.  
For both Morse potential and power law potential, 
rebinding is important for weaker bonds.  
The weaker bonds not only have smaller upper limit for applying Kramers 
rate theory but also have  greater lower limit for neglecting rebinding, 
therefore the deviation of Eq.~(\ref{eq:fmax}) is more serious for weaker 
bonds.
The rupture force $F_{max}$ in the regime where rebinding is important 
can be calculated
numerically by extending cubic theory to include nonzero $k_{on}$ in 
Eq.~(\ref{eq:rates}) into Eq.~(\ref{eq:rupturetime}) 
and ~(\ref{eq:ruptureforce}).
The long dashed lines in Fig.~2 shows prediction of this 
{\em extended cubic theory}
in regions where rebinding is not negligible. 
Indeed extended cubic theory provides a good prediction for rupture force in 
regions where Eq.~(\ref{eq:fmax})
deviates from simulation result.    
   
It is interesting to notice that at low loading rate, there is still some
difference between the rupture force in {\em extended cubic theory} 
and simulation.   
This reveals the limit of theoretical description for DFS of 
simple bonds.  Figure 3 shows power law potential with $W=10$, the case where
the difference between extended cubic theory and simulation is most 
significant, and the potential predicted by cubic theory.  
Since the cubic theory potential is an expansion of $U_0(x)$ near $x_1$, 
it does not fit $U_0(x)$ well for $x$ smaller than $L_0$.  
Thus {\em extended cubic theory} does not have high accuracy 
when bond rupture time $t$ predicted by extended cubic theory
has $x_a(t) < L_0$, and that is when
\begin{eqnarray}
  F_{max} < U_0'(x_1) - \frac{c}{2}(x_1-L_0)^2,
\label{eq:cubiclimit}
\end{eqnarray}
where $F_{max}$ is calculated from the {\em extended cubic theory}.  
The cross marks in Fig.~2 are where 
$ F_{max} = U_0'(x_1) - \frac{c}{2}(x_1-L_0)^2$, indeed they mark the
limiting loading rates for high precision prediction of  
extended cubic theory.  Finally, we note that in analyzing experimental data,
the fitting to extended cubic theory uses data both inside and outside the 
high precision region. Therefore the resulting information like binding energy
and $k_{on}$ should be very close to true values.  
This explains why data fitting in Ref.~\cite{ref:Dudko_PNAS_03} is so 
successful for Morse potential.  Thus, extended cubic theory, an attempt to 
mimic the potential landscape of a bond near where bond rupture occurs, 
not only captures the basic physical picture of bond rupture, but also provides
good quantitative result for analyzing experimental data.
 
  In summary, we have studied the DFS of a simple soft bond theoretically.
Our analysis shows that rebinding is important at low loading rate, 
thermal effect is negligible at high loading rate, in the intermediate loading
rate the rupture force scales as $const. + (\ln (kv))^{2/3}$.  
The numerical simulations for a couple of model potentials with typical 
binding energy show that the prediction of scaling law is 
better for bonds with stronger binding energy.  
Rebinding is important at low loading rate, and 
{\em extended cubic theory} fixes the deviation between Eq.(~\ref{eq:fmax}) 
and simulation. 
Our study shows that {\em extended cubic theory} provides the basic physical 
picture of the rupture of simple soft bonds.
It would very interesting to extend the present model to multiple bonds 
in series and multiple bonds in parallel 
and compare to the analysis done in previous 
studies~\cite{ref:Hummer_BPJ_03,ref:Seifert_EPL_02}, 
and that is our future direction.

We would like to thank Prof. David Lu for stimulating discussion in the early
stage of this work.  This work is supported by the National Science Council of the Republic of China (Taiwan) under grant no. NSC 92-2112-M-008-019.

\newpage
\noindent
{\large Figure captions}\\
\begin{itemize}
\item Figure 1: The solid curve is $\tilde{F}(x)$. 
When $t_1 < t < t_2$ there are three solutions to the
equation $-(\partial U_0/\partial x) - k(x-L_0)=-kvt$, 
they correspond to the potential wells and the barrier between them.
\item Figure 2: (a). Rupture force for Morse potential 
$U_0(x)=W ([1-\exp (-2b(x-1))]^2-1)$ with $b=1.5$
from both numerical simulation and Eq.~(\ref{eq:fmax}).
(b). Rupture force for a power
law potential $U_0(x) = W(\frac{1}{2x^6}-\frac{3}{2x^2})$ from both
numerical simulation and  Eq.~(\ref{eq:fmax}).
Simulation result for $W=45$ are presented as triangles, 
circles $W=30$, diamonds $W=10$. 
Filled symbols in the high $v$ region are where Kramers rate theory 
cannot describe bond rupture, filled symbols in the low $v$ region are
are where rebinding is important.  Thick dashed lines are from
{\em extended cubic theory}, other lines are from Eq.~(\ref{eq:fmax}).
Cross marks are where $F_{max} = U_0'(x_1) - \frac{c}{2}(x_1-L_0)^2$,
i.e.,
{\em extended cubic theory} does not predict $F_{max}$ with high precision
when $v$ is smallar than the cross marks. There is no cross mark for 
Morse potential with $W=45$, in this case cubic theory agrees with 
simulation with high precision within the range of our simulation.
\item Figure 3: Solid line: power law potential with $W=10$, dashed line: 
potential predicted by cubic theory.  The prediction of 
{\em extended cubic theory} is not highly accurate when 
$x_a(\langle t_r \rangle)$ is located at region where cubic theory
does not fit $U_0(x)$ well, i.e., when 
$x_a(\langle t_r \rangle)<L_0=1$.  
\end{itemize}

\end{document}